# An algorithm to estimate unsteady and quasi-steady pressure fields from velocity field measurements

John O. Dabiri[1*], Sanjeeb Bose[2], Brad J. Gemmell[3], Sean P. Colin[4], and John H. Costello[5]

[1]*Graduate Aeronautical Laboratories and Bioengineering, California Institute of Technology, Pasadena, California 91125, USA*

[2]*Center for Turbulence Research, Stanford University, Stanford, California 94305, USA*

[3]*Whitman Center, Marine Biological Laboratory, Woods Hole, Massachusetts 02543, USA*

[4]*Marine Biology and Environmental Sciences, Roger Williams University, Bristol, Rhode Island 02809, USA*

[5]*Biology Department, Providence College, Providence, Rhode Island 02918, USA*

*Author for correspondence (email: jodabiri@caltech.edu)

Short title: Pressure estimation from velocity fields

**Summary**

We describe and characterize a method for estimating the pressure field corresponding to velocity field measurements, such as those obtained by using particle image velocimetry. The pressure gradient is estimated from a time series of velocity fields for unsteady calculations or from a single velocity field for quasi-steady calculations. The corresponding pressure field is determined based on median polling of several integration paths through the pressure gradient field in order to reduce the effect of measurement errors that accumulate along individual integration paths. Integration paths are restricted to the nodes of the measured velocity field, thereby eliminating the need for measurement interpolation during this step and significantly reducing the computational cost of the algorithm relative to previous approaches. The method is validated by using numerically-simulated flow past a stationary, two-dimensional bluff body and a computational model of a three-dimensional, self-propelled anguilliform swimmer to study the effects of spatial and temporal resolution, domain size, signal-to-noise ratio, and out of plane effects. Particle image velocimetry measurements of a freely-swimming jellyfish medusa and a freely-swimming lamprey are analyzed using the method to demonstrate the efficacy of the approach when applied to empirical data.

## Introduction

A longstanding challenge for empirical observations of fluid flow is the inability to directly access the instantaneous pressure field using techniques analogous to those established to measure the velocity field. Recent approaches have made significant progress, especially in the measurement of pressure associated with unsteady fluid-structure interactions (e.g. Hong and Altman, 2008; Jardin et al. 2009a, 2009b; David et al., 2009; Rival et al. 2010a, 2010b; David et al., 2012; Tronchin et al., 2012; van Oudheusden, 2013; Liu and Katz, 2013). However, prior efforts have not achieved explicit pressure estimation for moving bodies with time-dependent shape, such as those characteristic of animal locomotion and feeding. The pressure field of swimming animals is complicated by the interaction between pressure associated with vortices in the flow and the irrotational pressure field due to acceleration of the body, often referred to as the acceleration reaction or added mass (Daniel, 1984).

Existing methods for empirical pressure estimation often require relatively complex measurement techniques such as multi-camera or time-staggered, multi-exposure particle image velocimetry (Jensen and Pedersen, 2004; Liu and Katz, 2006). In addition, significant computational costs can be associated with the post-processing required to derive the pressure field from measurements of the velocity or acceleration fields. These post-processing approaches generally fall into one of two categories. In the first case, the pressure field is computed as a solution to a Poisson equation, e.g. in an inviscid flow:

$$\nabla^2 p = -\rho\left(\nabla \cdot \frac{D\mathbf{u}}{Dt}\right), \tag{1}$$

where $p$ is the pressure, $\mathbf{u}$ is the velocity vector, $\rho$ is the fluid density, and $D/Dt$ is the material derivative, i.e., the time rate of change of an idealized infinitesimal fluid particle in the flow. Solution of equation (1) poses challenges in practice because measurement errors accumulate due to the required temporal and spatial derivatives of $\mathbf{u}$, the condition number (i.e. sensitivity) of the Laplacian operator (Golub and Van Loan, 1996), and measurement uncertainty in the boundary conditions, especially at fluid-solid interfaces (Gurka et al., 1999). For attached flows at high Reynolds number, the Neumann boundary condition specifying the pressure gradient at fluid-solid interfaces is given by the boundary layer approximation as $\partial p/\partial n \approx 0$, where $n$ is the direction of the local normal surface vector (Rosenhead, 1963). However, for separated flows at moderate or low Reynolds numbers, such as those commonly found in animal locomotion, *a priori* determination of the appropriate fluid-solid boundary conditions for solution of (1) can be intractable.

A second category of approaches for pressure field estimation is those based on direct integration of the pressure gradient term in the Navier-Stokes equation, e.g. for incompressible flow:

$$\nabla p = -\rho\left(\frac{D\mathbf{u}}{Dt} - \nu\nabla^2\mathbf{u}\right) \tag{2}$$

where $\nu$ is the kinematic viscosity of the fluid. The pressure difference between two points in the domain is determined by integration of equation (2) between the two points. For example, the difference in pressure between two points $\mathbf{x}_1$ and $\mathbf{x}_2$ is given by

$$p_2 - p_1 = \int_{\mathbf{x}_1}^{\mathbf{x}_2} \nabla p d\mathbf{x} \qquad (3)$$

Because measurement errors accumulate along the path of integration from $\mathbf{x}_1$ to $\mathbf{x}_2$ in equation (3), various techniques have been employed to make this approach less sensitive to measurement uncertainty. A common strategy is to take advantage of the scalar property of the pressure field, such that its local value is independent of integration path. Therefore, each independent integration path that arrives at a point in the flow is in principle an independent estimate of the pressure at that point, provided that measurement errors are uncorrelated. By polling a large number of integration paths, an estimate of the local pressure can be achieved. For example, one successful method (Liu and Katz, 2006) uses an iterative scheme that averages $2m(n+m) + 2n(2m+n)$ integration paths on an $m$ x $n$ grid in order to estimate the instantaneous pressure field.

The aforementioned iterative scheme, while effective in limiting the influence of measurement errors, still incurs a relatively high computational cost. For example, for a 128 x 128 grid of velocity vectors that is commonly acquired using PIV, the method requires $1.6 \times 10^5$ integration paths per iteration of velocity field integration; and several iterations can be required for convergence of the method (Liu and Katz, 2006). Furthermore, if each integration path is taken as a straight line through the domain, then the method requires interpolation of the estimated pressure gradient field in order to evaluate integration path points that do not coincide with the original data grid. While these requirements are not necessarily prohibitive for two-dimensional calculations, they are time-consuming and are indeed a showstopper for extension of the method to three dimensions.

We present a simple yet demonstrably effective approach for pressure estimation that is in the spirit of the second category of pressure estimation methods. The method is validated by using two numerically-simulated flows: flow past a two-dimensional, stationary bluff body and the flow created by a three-dimensional, self-propelled anguilliform swimmer. The first flow is used to characterize a quasi-steady implementation of the algorithm, in which the pressure field is estimated from a single velocity field measurement. The second flow demonstrates the ability of the method to accurately estimate the pressure on unsteady, deformable bodies such as those of relevance in animal locomotion. Both flows are used to characterize the method, including its numerical convergence properties and sensitivity to domain size, signal-to-noise ratio, and out of plane effects. Furthermore, we apply the method to PIV measurements of a freely-swimming jellyfish medusa and a freely-swimming lamprey, showing that this tool can be applied to the type of measurement data commonly acquired in research.

## Materials and Methods

*Material acceleration estimation*

The instantaneous fluid particle acceleration $D\mathbf{u}/Dt$ required for calculation of the pressure gradient in equation (2) is estimated by advecting idealized infinitesimal fluid particles in the measured velocity fields. For quasi-steady estimation, the material acceleration is derived from a single velocity field as

$$\frac{D\mathbf{u}}{Dt}(\mathbf{x}_i) \approx \frac{\mathbf{u}_{PIV}(\mathbf{x}_i^{\,a}) - \mathbf{u}_{PIV}(\mathbf{x}_i)}{\Delta t} \qquad (4)$$

where $i = 1, 2... m$ x $n$ (i.e. the dimensions of the velocity grid), $\mathbf{x}_i$ are the positions of fluid particles coincident with the grid points in the PIV velocity field, and $\mathbf{x}_i^a$ are the positions of those fluid particles after being advected by the instantaneous velocity field for a period $\Delta t$:

$$\mathbf{x}_i^{\,a} \approx \mathbf{x}_i + \mathbf{u}_{PIV}(\mathbf{x}_i)\Delta t \qquad (5)$$

In order for equations (4) and (5) to remain valid, $\Delta t$ is limited to values much smaller than the characteristic time scale of the flow, yet sufficiently large that there is a measurable change in the fluid particle velocity.

For many flows, especially those involving accelerating or deforming bodies, the aforementioned constraint on $\Delta t$ cannot be satisfied. For these inherently unsteady fluid-structure interactions, we derive the material acceleration from two sequential velocity fields as

$$\frac{D\mathbf{u}}{Dt}(\mathbf{x}_i, t_1) \approx \frac{\mathbf{u}_{PIV}(\mathbf{x}_i{}^a, t_2) - \mathbf{u}_{PIV}(\mathbf{x}_i, t_1)}{t_2 - t_1} \tag{6}$$

where $\mathbf{x}_i{}^a \approx \mathbf{x}_i + \left( \frac{\mathbf{u}_{PIV}(\mathbf{x}_i, t_1) + \mathbf{u}_{PIV}(\mathbf{x}_i, t_2)}{2} \right)(t_2 - t_1)$ (7)

Equation (7) is akin to a Crank-Nicolson (i.e. trapezoidal) scheme for the particle positions, in contrast to the forward Euler scheme in equation (5). Hence, the convergence of the method with time step is second order (Crank and Nicolson, 1947).

The primary source of measurement error in this type of unsteady estimate of the material acceleration $D\mathbf{u}/Dt$ arises from temporal noise in the measured velocity components at each node in the velocity field. We address this by applying a temporal filter to the time series of velocity fields, which results in a smoothing spline approximation $\mathbf{u}^*$ to the velocity $\mathbf{u}$ at each node in the velocity field. The spline approximations are defined such that they minimize, for each component of $\mathbf{u}$, the parameter

$$S_{\mathbf{u}} = \phi \sum_{\tau=1}^{N} (\mathbf{u}_\tau - \mathbf{u}_\tau{}^*)^2 + (1-\phi) \int_{t_{\min}}^{t_{\max}} \left( \frac{d^2\mathbf{u}^*}{dt^2} \right)^2 dt \tag{8}$$

where $\tau = 1...N$ is the temporal sequence of velocity fields to be analyzed, $\mathbf{u}_\tau$ is a velocity vector corresponding to velocity field $\tau$ in the sequence, $\mathbf{u}_\tau{}^*$ is the spline-approximated value of the same velocity vector for the same velocity field in the sequence, $t_{\min}$ and $t_{\max}$ are the temporal bounds on the sequence of velocity fields, and $\phi$ is a weight between the first and second terms and has a value between 0 and 1. In effect, the parameter $S_{\mathbf{u}}$ quantifies both the deviation of the spline approximation from the original data (i.e. the first term) and the total curvature magnitude of the spline approximation (i.e. the second term). For $\phi = 0$, only the second term is minimized, resulting in a least-squares fit with zero curvature, i.e. a linear fit to the data. For $\phi = 1$, only the first term is minimized, giving a cubic spline fit that passes through each original data point. In all that follows, we set $\phi = 0.05$, a value we have identified as enabling effective temporal noise filtering without discarding true temporal trends in the measurement data.

Further characterization of the temporal filter is provided in Appendix 2. In particular, it is shown that the use of the temporal filter increases the order of temporal convergence above second order, as anticipated by theory (Atkinson, 1968).

It is worth noting that the distinction between the quasi-steady and unsteady approaches can be made explicit by decomposing the material acceleration into its Eulerian components:

$$\frac{D\mathbf{u}}{Dt} \equiv \frac{\partial \mathbf{u}}{\partial t} + (\mathbf{u} \cdot \nabla)\mathbf{u} \qquad (9)$$

The quasi-steady approximation in equations (4) and (5) implicitly neglects the first term on the right-hand side of equation (9), whereas the unsteady calculation retains it.

The viscous term on the right-hand side of equation (2) is computed using centered finite differences between adjacent nodes in the velocity field. The effect of the viscous term is evaluated in the context of a numerical simulation described in the validation section.

*Pressure gradient integration*

Whereas previous methods that integrate the pressure gradient via many integration paths assign to each grid point the arithmetic mean of the many integrations, in the present approach the paths are polled by taking the median. The median is less sensitive to grossly erroneous values that may arise on a few of the integration paths due to localized measurement errors or due to localized errors created by the aforementioned material acceleration approximations in equations (4) through (7). Hence, this approach enables a significant reduction in the total number of integration paths per frame that are required to achieve accurate pressure estimates. Fig. 1 illustrates the paths used presently. Eight families of integration paths are used, with each family originating at the domain boundary and propagating toward each grid point from the left (L), upper left (UL), top (T), upper right (UR), right (R), lower right (LR), bottom (B), and lower left (LL), respectively.

Only 8 integration paths (one per family) per grid point are used, for a total of $8m \times n$ paths per velocity field. For the aforementioned example grid of 128 x 128 velocity vectors, $1.3 \times 10^5$ integration paths are required, a 20 percent reduction from existing optimal methods (Liu and Katz, 2006). More importantly, the integration paths are constrained to include only grid points coincident with the original velocity field. For example, the UL integration path is comprised of alternating integration steps in the -y and +x directions, originating at the domain boundary and terminating at each grid point. Hence, no interpolation is required in order to integrate the pressure gradient field. Furthermore, portions of many of the paths are redundant, facilitating fast calculation using simple matrix manipulations. A forward Euler spatial integration scheme is used throughout, resulting in first-order spatial convergence of the method (see Appendix 1).

An important limitation of the present algorithm that arises from the trade-off between speed and accuracy is that it assumes the pressure is zero at the point on the outer domain boundary where each integration path is initiated. This does not imply, however, that the final pressure estimate is constrained to be zero at the boundaries. Integration paths that originate from the other domain boundaries and terminate at a given boundary may estimate a non-zero value of pressure at the termination point. If the median of all paths terminating at that point on the domain boundary is non-zero, then the final pressure estimate at that point will also be non-zero. Note that for all points in the domain, the final pressure estimate is relative to a zero reference pressure, as that is the pressure at the origin of each integration path. The impact of these assumptions on the robustness of the technique is quantified below, and it is shown to be modest for the external flows tested. At the same time, the net result of this tradeoff in the algorithm design is a more than order-of-magnitude reduction in computational time compared to previous methods (see Appendix 2).

A common source of localized error that can affect pressure estimates is the presence of solid objects in the flow. Typical PIV measurements are often unreliable in the region close to solid objects, which compromises pressure integration paths that cross the fluid-solid interface, especially in previous methods that average the erroneous data instead of discarding it via median polling (or in Poisson solvers that rely on the pressure gradient at the fluid-solid interface as a boundary condition). In the present algorithm, integration paths that cross a fluid-solid interface in the flow can be nullified by assigning the nodes nearest to the interface an undefined pressure gradient. Hence, when that value is integrated along any integration path, the pressure value for that path also becomes undefined and therefore does not contribute to the median calculation.

*Validation data sets*

To validate the accuracy of the quasi-steady pressure estimates achieved using this algorithm, a numerical simulation of flow past a two-dimensional square cross-section cylinder at a Reynolds number of 100 was used. This numerical data set enabled quantification of the effects of spatial resolution, domain size, and signal-to-noise ratio, while providing a known pressure field standard for comparison (see Appendix 1). The numerical simulation was executed using a solver that computes on arbitrary polyhedra (Ham and Iaccarino, 2004). In the present case, a regular Cartesian mesh was utilized and subsequently interpolated onto coarser grids of varying sizes typical of PIV data. The viscous term in equation (2) was retained in all of the calculations to demonstrate the robustness of the median polling approach to errors normally associated with application of the Laplacian operator. For all calculations of equations (4) and (5) in this validation, we set $\Delta t = 0.01h/U_{max}$, where $h$ is the mean grid spacing and $U_{max}$ is the maximum flow speed in the measurement domain. The results described below were insensitive to order of magnitude larger and smaller values of $\Delta t$. Where noted, nearest-neighbor Gaussian smoothing was applied both to the pressure gradient before integration and to the resulting pressure field.

The accuracy of the fully unsteady pressure estimates was validated by using a published numerical simulation of a three-dimensional, self-propelled anguilliform swimmer (Kern and Koumoutsakos, 2006). The Reynolds number based on swimmer length and speed was approximately 2400. Time steps between sequential velocity fields from $0.02T$ to $0.08T$ (where $T$ is the swimming stroke duration) were studied to quantify the temporal convergence of the method. The validation results described below are based on calculations of equations (6) and (7) using velocity fields separated by $0.02T$.

*Empirical data sets*

The present method was also applied to particle image velocimetry measurements of a freely-swimming *Aurelia aurita* (Linnaeus, 1758) jellyfish medusa and a freely-swimming *Anguilla rostrata* (LeSueur, 1817) lamprey to demonstrate the performance of the algorithm with empirical data inputs and, in the case of the jellyfish, without treatment of fluid-solid interfaces. The swimming Reynolds numbers of the jellyfish and lamprey were approximately 1000 and 10,000, respectively, and the time between sequential velocity fields was 5 ms ($t/T \approx 0.013$) and 4 ms ($t/T \approx 0.015$), respectively. Details of the PIV implementation can be found in published literature (Colin et al., 2012).

## Results and discussion

### *Quasi-steady pressure estimation*

Fig. 2 compares an instantaneous pressure field from the numerical simulation of flow past a stationary bluff body with the pressure field estimated from the corresponding velocity field using the quasi-steady form of the present algorithm. A vector field spatial resolution of $D/16$ (where $D$ is the side length of the bluff body) is used in the horizontal and vertical directions to mimic typical PIV measurements. The salient features of the flow, especially the high pressure on the upstream face of the bluff body and the low pressure in the shear layers and near-wake vortices, are well captured by the algorithm (see Appendix 1 for discussion of discrepancies in the far wake). Furthermore, the maximum and minimum pressures in the field are in quantitative agreement (Fig. A1). To be sure, nearest-neighbor Gaussian smoothing creates a spurious thin layer of undefined pressure at the fluid-solid interface and moves the pressure peak on the upstream face of the body away from the interface. This artifact can be corrected by application of a more sophisticated smoothing filter near fluid-solid interfaces. However, for the present proof-of-concept demonstration, the correct near-body pressure is recovered by increasing the grid resolution so that the nearest-neighbor filter artifact on the body surface is limited to a smaller region very close to the body. Additional surface pressure calculations for the quasi-steady case (Appendix 1) are based on a velocity vector spacing of $D/64$. Note that a similar increase in resolution using a PIV camera would require a concomitant reduction in the measurement window size due to limits on camera pixel resolution.

Additional characterization of the quasi-steady algorithm is detailed in Appendix 1, including analysis of spatial convergence; the relative contribution of each integration path to the median pressure field; robustness to measurement noise; and the effects of domain size, fluid viscosity, and fluid-solid interfaces.

### *Unsteady pressure estimation*

Fig. 3 compares an instantaneous pressure field from the numerical simulation of a self-propelled anguilliform swimmer with the pressure field estimated from the corresponding velocity field using the unsteady form of the present algorithm. A vector field spatial resolution of $L/42$ (where $L$ is the length of the swimmer) is used in the horizontal and vertical directions. No smoothing is applied to this data set in order to contrast the results with those in the previous section. The algorithm is effective in capturing the high-low pressure couples formed on the sides of the swimmer head and tail as they accelerate in the positive-$y$ direction; the low-high pressure couple formed at the mid-body as it accelerates in the negative-$y$ direction; and the pressure in the wake vortices.

The importance of the unsteady term in equation (9) is illustrated by comparison with the pressure field estimated using the quasi-steady approximation, shown in Fig. 3C. Low pressure in the wake vortices is captured, but the high-low pressure couples on the body surface due to the body added mass are missing entirely, as is the high pressure in the wake due to vortex added mass (Dabiri, 2006). The comparison is further quantified in Fig. 4, which plots the pressure on a contour surrounding the swimmer and immediately adjacent to the region of undefined pressure. At each of the four phases of the swimming cycle shown, good agreement is achieved between the pressure computed in the numerical simulation and the pressure estimated from the velocity field using the unsteady algorithm. By contrast, the pressure estimated by the quasi-steady algorithm is erroneous everywhere except near the forming wake vortex at the tail.

Additional characterization of the unsteady algorithm is provided in Appendix 2, including analysis of temporal convergence and out-of-plane effects for three-dimensional flows.

To demonstrate the efficacy of the present method for analyzing empirically measured velocity fields, Fig. 5 shows measured velocity and vorticity fields for the freely-swimming jellyfish and lamprey (panels A and C) along with the corresponding pressure fields estimated using the unsteady algorithm (panels B and D). The full measurement domain is shown in both cases; the velocity vector field is plotted at half of the full resolution. Only the left half of the jellyfish body is visible in the measurement domain; its exumbrellar surface is indicated by a black curve in panels A and B. The full lamprey body is visible in panels C and D.

In both cases, the pressure field derived from the velocity field measurements captures key features near the body surface and in the wake. In particular, the jellyfish data set indicates low pressure in the forming starting vortex and high pressure where the bell margin is accelerating inward and pushing the adjacent fluid. The results are consistent with the measured vorticity field (panel C), with the region of low pressure corresponding to the core of the starting vortex. The presence of low and high pressure regions near the bell margin is also in agreement with previous numerical simulations of a swimming jellyfish with similar body shape and kinematics (Sahin et al., 2009).

The lamprey data set shares similarities with the three-dimensional numerical model shown previously. The vorticity and pressure fields are less smooth and show finer structure in the empirical measurements, which is attributable in part to the Reynolds number being approximately four times higher than that of the numerical simulation.

The ease of implementation of this algorithm, both in terms of data acquisition and velocity field post-processing, and its relatively low computation cost (see Appendix 2) gives it the potential to find use in a broad range of problems of interest in biological fluid mechanics. Because the temporal filter implemented in the unsteady algorithm does add considerable time to the pressure calculation (cf. Fig. A10), in practice one should first evaluate the results of both the quasi-steady and the fully unsteady implementations of the algorithm on a sample of the data of interest to determine whether unsteady effects are important. If they are not, then the quasi-steady calculation provides the most efficient tool for determination of the pressure field.

Although the present evaluation focused on two-dimensional velocity fields, it is straightforward to extend the algorithm to three dimensions by the addition of a limited number of new integration paths consistent with the geometry in Fig. 1. In that case, even greater reductions in computation expense can be achieved relative to existing methods due to the relatively small total number of required integration paths and the elimination of associated velocity field interpolation during integration of the pressure gradients.

A free MATLAB implementation of this algorithm is available at http://dabiri.caltech.edu/software.html.

## Appendix 1: Additional Characterization of Quasi-steady Algorithm

*Effect of median polling*

To illustrate the contribution of each integration path to the final pressure estimate, Fig. A1 plots the pressure on the body surface (at 0.1$D$ away from the fluid-solid interface, to avoid the spatial filter artifact) and on two additional concentric square contours in the domain (e.g. the dashed contour in Fig. A2A), as computed using each of the 8 families of integration paths. The results illustrate the benefit of median polling versus an average of the integration path results.

For example, only 5 of the integration path families intersect the upstream face of the bluff body without passing through the body itself. The median of these curves is in good quantitative agreement with the correct surface pressure (Fig. A1A). The pressure profiles for the two concentric square contours in the domain (i.e. Figs. A1B and A1C) indicate that the contribution of each family to the final pressure estimate is spatially non-uniform. This is illustrated qualitatively in Fig. A1D, which is a contour plot that colors each point in the domain according to the path family that corresponds to the median pressure at that point. Because there are 8 path families, the median is always the average of the two intermediate values (where none of the paths is undefined due to intersection with the solid body). To reveal the individual integration path family contributions, a ninth pressure value equal to the mean of the 8 path families is included in Fig. A1D, so that the median pressure is from either a single integration path family or from the mean. The contour plot indicates that each integration path family contributes to the final pressure field estimate, but the contributions are often spatially localized. The pressure estimates for the R family of integration paths are noticeably less accurate than the other families (e.g. Fig. A1A) and yet, as illustrated in Fig. A1D, these paths determine the pressure estimate in the far wake. This leads to the observed poorer pressure estimate in that region of the flow (e.g. Fig. 2B). The underlying source of this effect is discussed below in the section examining the effect of boundary conditions.

*Effect of global measurement error*

Perhaps the most important test of the algorithm is its robustness to global measurement errors, such as those associated with empirical measurements. Fig. A2 illustrates the streamwise velocity contours for data sets with increasing levels of Gaussian white noise superimposed on the $u$ and $v$ velocity components. The highest levels of noise, corresponding to the lowest signal-to-noise ratios, are higher than typical PIV data but possibly representative of instantaneous two-dimensional data collected in a highly-turbulent flow field, where out-of-plane motion can reduce data quality. Comparison of the pressure profiles on a square contour centered on the bluff body and with side length $3D$ so that it passes through the salient flow features (i.e. Fig. A2A) indicates that, with the exception of the highest noise level tested, the quantitative pressure estimates remain consistent with the noise-free result despite relatively high noise (Fig. A3A).

Error in the pressure estimate is not additive with the increasing noise level because errors do not accumulate uniformly on the 8 paths that arrive at each point in the domain. Hence, median polling remains an effective filter irrespective of the noise magnitude, up to the second-highest noise level tested. At higher noise levels, contour plots of the pressure estimate begin to exhibit spatial discontinuities reminiscent of the median contributions in Fig. A1D. Because the pressure estimates from each integration path family begin to diverge in the presence of high noise levels, median polling in this case leads to spatially discontinuous changes in pressure. Result of this sort are an indication that measurement noise in the input velocity data has become unacceptably large.

*Effect of boundary conditions*

As mentioned previously, a major assumption implicit in the present algorithm is that the pressure on each integration path is zero at its originating point on the boundary, to avoid the need for a computationally expensive iteration scheme to solve for the boundary pressure as part of the field solution (Liu and Katz, 2006). Although this assumption can be reasonable for large domains, it is prudent to investigate the dependence of the pressure estimate on the domain size. Fig. A3B plots the pressure on a square contour centered on the bluff body (see Fig. A2A) for domains ranging in size from $H/D = 2$ to 30, where $H$ is the half-width of the domain. The results indicate that the accuracy of the algorithm (and hence, the assumption regarding the boundary pressure) is not significantly compromised until the domain shrinks to $H/D = 2$. This limitation is important to keep in mind when designing experiments that will make use of the present algorithm.

Notwithstanding the demonstrated efficacy of the aforementioned assumption regarding the boundary pressure, examination of the individual pressure estimates on each family of integration paths reveals that some individual estimates are severely compromised by this assumption. Most notably, the R family of integration paths originate at the downstream boundary of the domain, where vortices shed by the bluff body exit the measurement window and create a non-zero pressure on that boundary. Hence this family of pressure estimates is significantly less accurate than the others, as seen in Fig. A1A for example. The benefit of the median polling approach is that this estimate is usually discarded in determining the final pressure estimate. In contrast, previous methods would include pressure estimates affected by the downstream boundary in the final averaged pressure estimate, and therefore require additional computational effort to resolve the correct pressure on that boundary via iterative processes.

However, the present method does suffer in that the pressure in regions close to the downstream boundary is based either on integration paths that originate at the downstream boundary where the pressure is nonzero (i.e. R, UR, and LR families) or on long integration paths from the other boundaries. The relatively large error accumulated on the long integration paths can make them an even poorer estimate of the local pressure near the R boundary (cf. Fig. A1D); hence the median pressure in this region is less accurate than in the rest of the domain. This limitation is inherent in the present method and should be kept in mind when using the technique for flows with large velocity gradients at any of the boundaries.

*Effect of fluid viscosity*

It is useful to examine the role of the viscous term in equation (2), as many previous pressure estimation methods neglect this term. Fig. A4A plots the pressure estimates on the body surface for each integration path family as in Fig. A1A, but for a pressure estimate that neglects the viscous term in the Navier-Stokes equation. The effect is most noticeable in integration paths orthogonal to the mean flow (i.e. T and B), especially near the upstream face of the bluff body. This result can be understood by considering the contributions to the pressure gradient from the streamwise and transverse material acceleration components, $Du/Dt$ and $Dv/Dt$, relative to the contributions from the Laplacian of the streamwise and transverse velocity components in the viscous term. As the flow approaches the upstream face of the bluff body, the material acceleration is dominated by streamwise fluid particle deceleration $Du/Dt$. However, the pressure computed on integration paths that are orthogonal to the mean flow (i.e. T and B) is independent of this term. Instead, on these paths the pressure depends to a significant degree on the local velocity curvature (i.e. second spatial derivative) as the flow is turned around the bluff body. This effect is captured in part by the Laplacian of the transverse velocity. Hence, its neglect leads to an underestimate of the pressure on those integration paths. The net effect of the neglected viscous term is minimal due to the median polling approach implemented presently, i.e., the T and B paths do not represent the median pressure estimate on the upstream face of the bluff body and are therefore not a factor in the final pressure estimate in that region of the flow.

*Effect of fluid-solid interfaces*

An aspect of the present algorithm that can be cumbersome is the treatment of the fluid-solid interface to eliminate the effect of integration paths that pass through solid objects in the flow. For example, for moving objects, this approach requires the identification of the fluid-solid interface in each data frame. To illustrate the effect of the interface treatment in the algorithm, Fig. A4B plots pressure estimates on the body surface, where the algorithm has been implemented without treating the fluid-solid interface. The accuracy of the algorithm is noticeably affected due to additional spurious pressure estimates on paths that cross the body. However, it is noteworthy that the final pressure estimate is still qualitatively consistent with the correct pressure field. It may therefore be acceptable to bypass the fluid-solid interface treatment where only a quantitative approximation of the pressure field is sought. The results of the analysis in Fig. 5B, which did not identify the boundary of the medusa as was done for the numerical data, suggest that suitable pressure estimates can still be achieved where treatment of the fluid-solid interfaces is not practical.

*Spatial convergence*

The spatial convergence of the quasi-steady algorithm was evaluated by computing the pressure on a square contour immediately adjacent to the region of undefined pressure on the bluff blody, and by integrating the pressure to compute the net force in the streamwise and lateral directions. Fig. A5 plots the fractional error in these calculations (using the pressure from the numerical simulation (CFD) as the true value, i.e. $\left|F_{CFD} - F_{estimate}\right| / F_{CFD}$ ) versus the grid resolution. At relatively large grid spacing, the log-log curve has a slope of 1, indicating the expected first-order spatial convergence of the method. As the grid spacing is further reduced, the error decreases more slowly. This effect is due to a combination of inherent model error and numerical round-off error. A convergence plot for calculations of the time-averaged streamwise force is included. Its departure from first order convergence at small grid spacing confirms that the quasi-steady approximation is not solely responsible for errors at small grid spacing. For grid spacing less than $D$/16, the error falls below 5 percent for the instantaneous pressure and approaches 10 percent for the time-averaged pressure.

## Appendix 2: Additional Characterization of Unsteady Algorithm

*Effect of temporal filter*

Fig. A6 plots the time series of *v* component data at two selected points in the jellyfish PIV data set. Despite the relatively smooth spatial distribution of velocity, as illustrated in Fig. 5A and Fig. A7A, there is non-trivial scatter in the temporal data at both spatial locations. Flow accelerations computed by using finite differences of adjacent velocity fields would be subject to large errors because the local slope varies considerably between adjacent pairs of velocity fields. A temporal filter of the data is therefore essential in this case. Fig. A6 indicates the corresponding smoothing splines that were fit to the data and subsequently used to compute the material acceleration. The splines capture the true transient behavior of the flow while eliminating the high-frequency noise. Comparison of Fig. A7A and A7B illustrates that the spatial distribution of velocity is relatively unaffected by the temporal filter. It is prudent to note that if a flow exhibits real, high-frequency oscillations in the velocity, e.g. in turbulence, it will be essential that the PIV measurements are of sufficiently high temporal resolution such that the smoothing spline does not discard those temporal trends. In those cases, it is important that the frequency of PIV measurements satisfies the Nyquist sampling criterion with respect to the time scale of turbulence fluctuations, while concurrently avoiding sampling at frequencies high enough to incur numerical round-off errors in the calculation (Beckwith et al., 2007).

*Effect of out-of-plane flow*

Given that two-dimensional PIV data represents a projection of three-dimensional flow, it is useful to characterize the impact of that limitation on the accuracy of the present methods. As in prior work (Stamhuis and Videler, 1995), Fig. A8 characterizes the out-of-plane motion by computing the divergence of a two-dimensional velocity field extracted from the three-dimensional numerical simulation of the self-propelled swimmer and of the PIV data sets examined in Fig. 5. The divergence is made dimensionless by multiplying it by the time step between adjacent velocity fields, as this time scale is most relevant for calculation of the material acceleration. The plots effectively quantify the fractional change in the volume of an idealized infinitesimal fluid particle between adjacent velocity fields. Because the flows are incompressible, the fractional volume change would be identically zero if the flows were two-dimensional. Deviation from zero values can therefore be attributed to velocity gradients perpendicular to the plane of the velocity field, i.e., out-of-plane flow.

The results in Fig. A8 indicate that the three-dimensional numerical simulation exhibits greater out-of-plane flow than the PIV measurements. Given the demonstrated accuracy of the algorithm in the case of the three-dimensional numerical data, we can conclude that the algorithm is robust to out-of-plane effects at the magnitudes found in typical PIV measurements. To be sure, the divergence metric does not capture out-of-plane flow where there is no flow gradient in the perpendicular direction. However, in such cases, the PIV would itself be difficult to acquire, as the seed particles would not remain in the plane of the laser sheet sufficiently long to enable temporal cross-correlation of their positions.

*Temporal convergence*

The temporal convergence of the unsteady algorithm was evaluated by plotting the fractional error in the pressure at the head of the self-propelled swimmer at an instant of high acceleration (using the pressure from the numerical simulation (CFD) as the true value, i.e. $\left|p_{CFD} - p_{estimate}\right| / p_{CFD}$ ) versus the time step between velocity fields (Fig. A9). Although the available data was limited to time steps from $0.02T$ to $0.08T$, the results are consistent with temporal convergence that is higher than second order, except as the smallest step size is approached, where further reduction in error is limited by inherent model error and numerical round-off error. At a temporal spacing of $0.02T$, the error in the pressure at the head is approximately 8 percent.

When the unsteady algorithm is applied to a sequence of velocity fields that are spaced too closely in time, leading to increased numerical error, the results appear similar to those described in the context of global measurement error (Appendix 1) in which the pressure contours exhibit spatial discontinuity reminiscent of Fig. A1D.

*Computational cost*

Fig. A10 plots the time required for a single 3-GHz processor to apply the temporal filter and to compute the pressure field for velocity fields from 32x32 to 256x256 nodes, which is a practical upper limit for typical PIV measurements due to camera pixel resolution. The time required for the temporal filter scales linearly with the number of nodes in the velocity field. The cost is independent of the number of velocity fields in the sequence of up to the tested several hundred frames of data. The computational cost of the subsequent pressure calculation scales sublinearly in the range tested, and it is significantly lower than the cost of the temporal filter in absolute terms. For example, for a 128x128 velocity field, each pressure field is computed in approximately 3 seconds, as compared to 46 seconds using an existing iterative algorithm (Liu and Katz, 2006).

**Acknowledgments**

The authors gratefully acknowledge S. Kern and P. Koumoutsakos for providing the three-dimensional numerical simulation of the self-propelled swimmer, and X. Liu for providing access to the pressure gradient integration code used in Liu and Katz (2006). This research was supported by Office of Naval Research awards N000140810918 and N000141010137 to J.O.D.

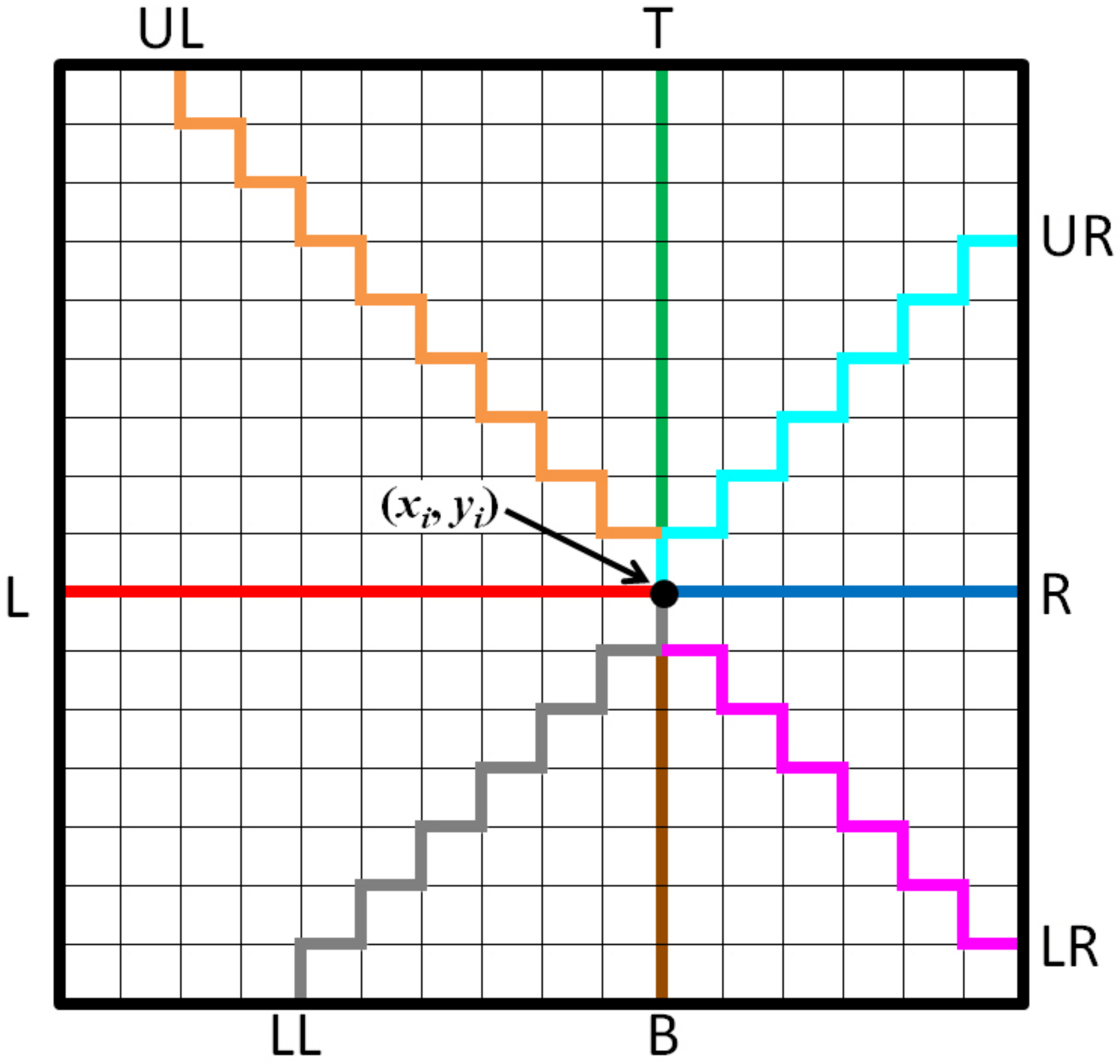

**Figure 1.** Geometry of integration paths for pressure field estimation. Eight paths originate from the domain boundary and propagate toward each point $(x_i,y_i)$ in the domain from the left (L), upper left (UL), top (T), upper right (UR), right (R), lower right (LR), bottom (B), and lower left (LL). The points on each path coincide with the measurement grid.

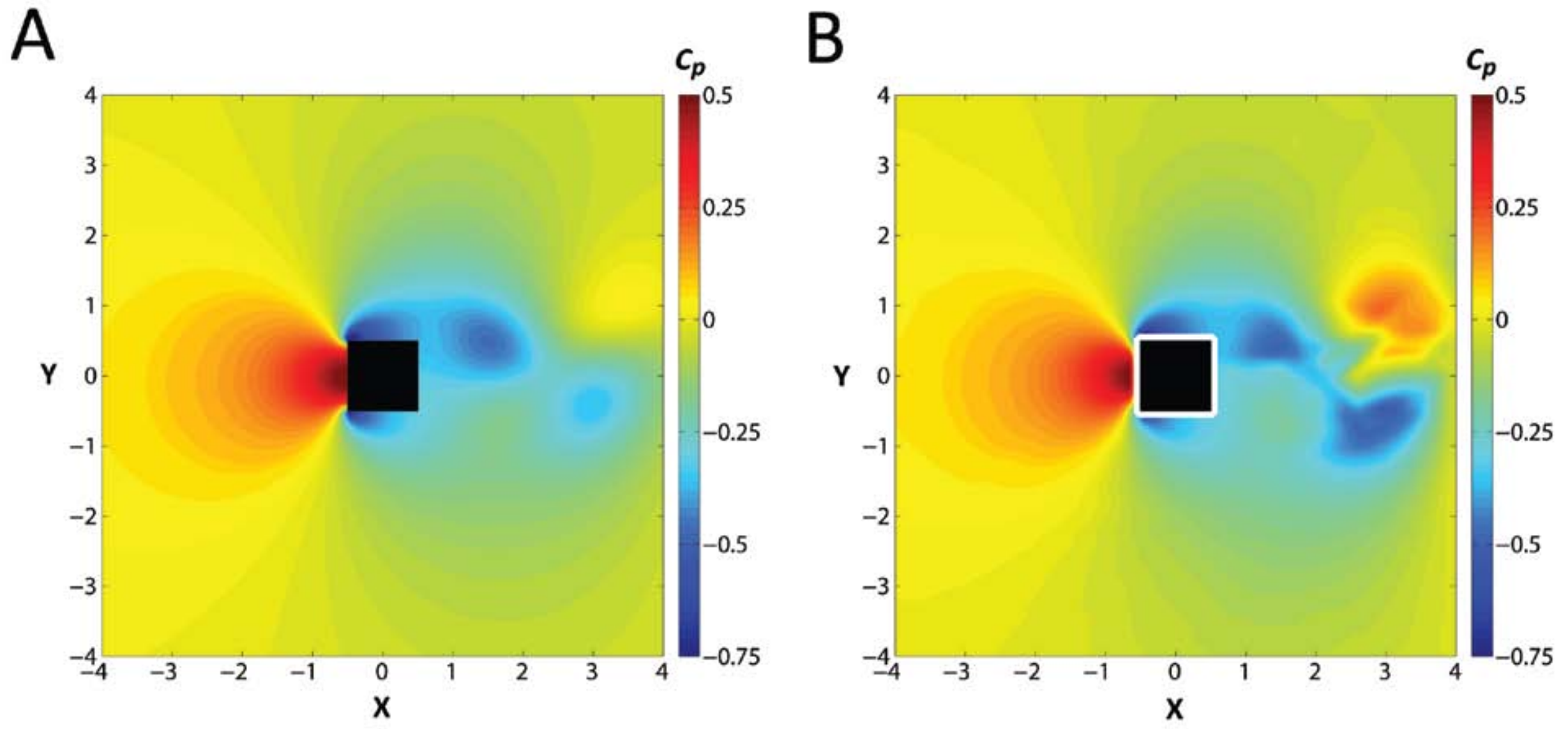

**Figure 2.** (A) Pressure field computed from numerical simulation of flow past two-dimensional square cylinder at a Reynolds number Re = $UD/\nu$ = 100. The pressure coefficient $c_p = p/(\rho U^2)$. (B) Pressure field estimated using quasi-steady algorithm.

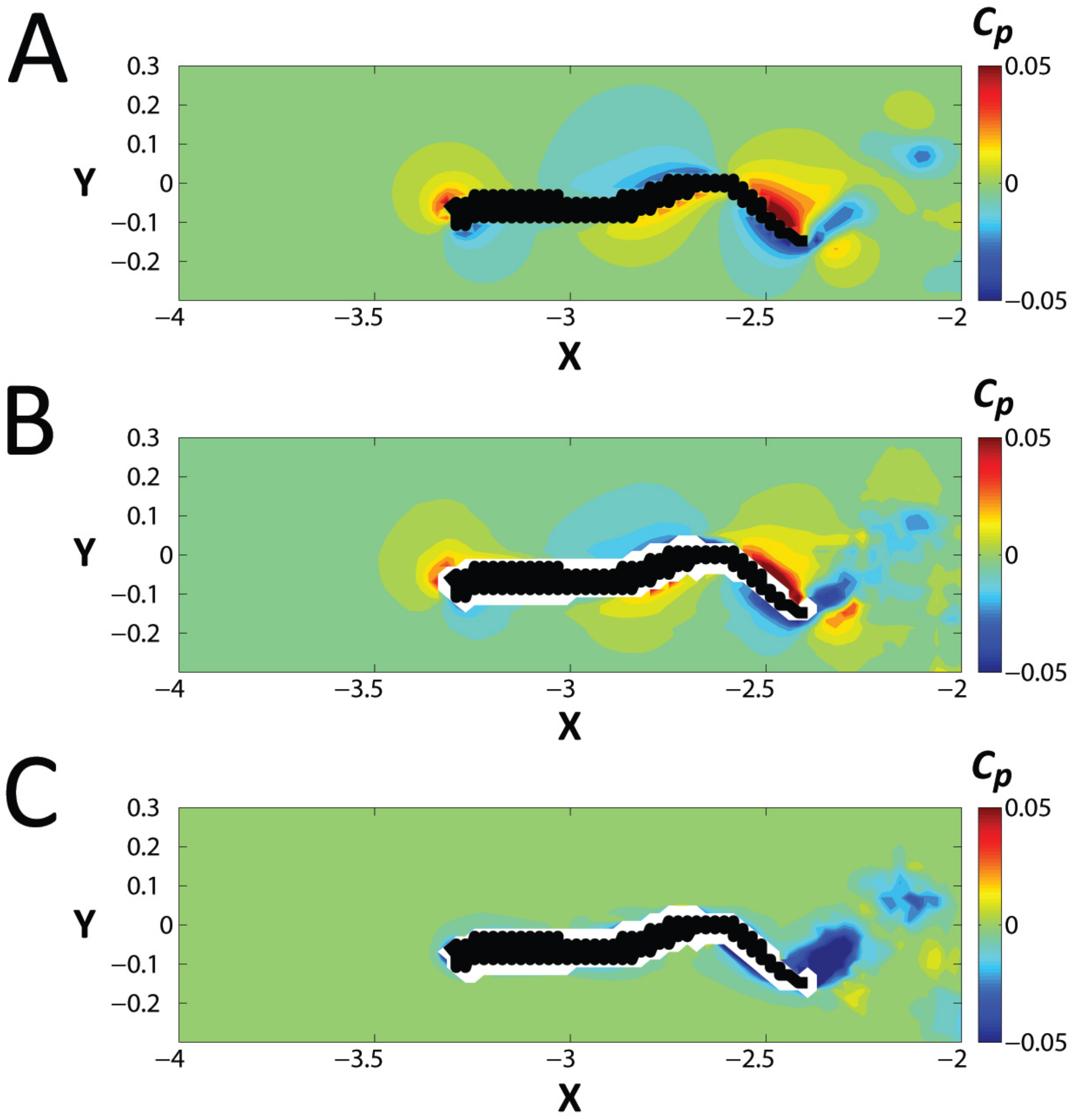

**Figure 3.** (A) Pressure field computed from numerical simulation of three-dimensional self-propelled swimmer. The pressure coefficient $c_p = p/(\rho U^2)$. Velocity nodes completely inside swimmer body are indicated in black (body surface is smooth in numerical simulation). Spatial coordinates are normalized by swimmer length. (B) Pressure field estimated using unsteady algorithm. (C) Pressure field estimated using quasi-steady algorithm.

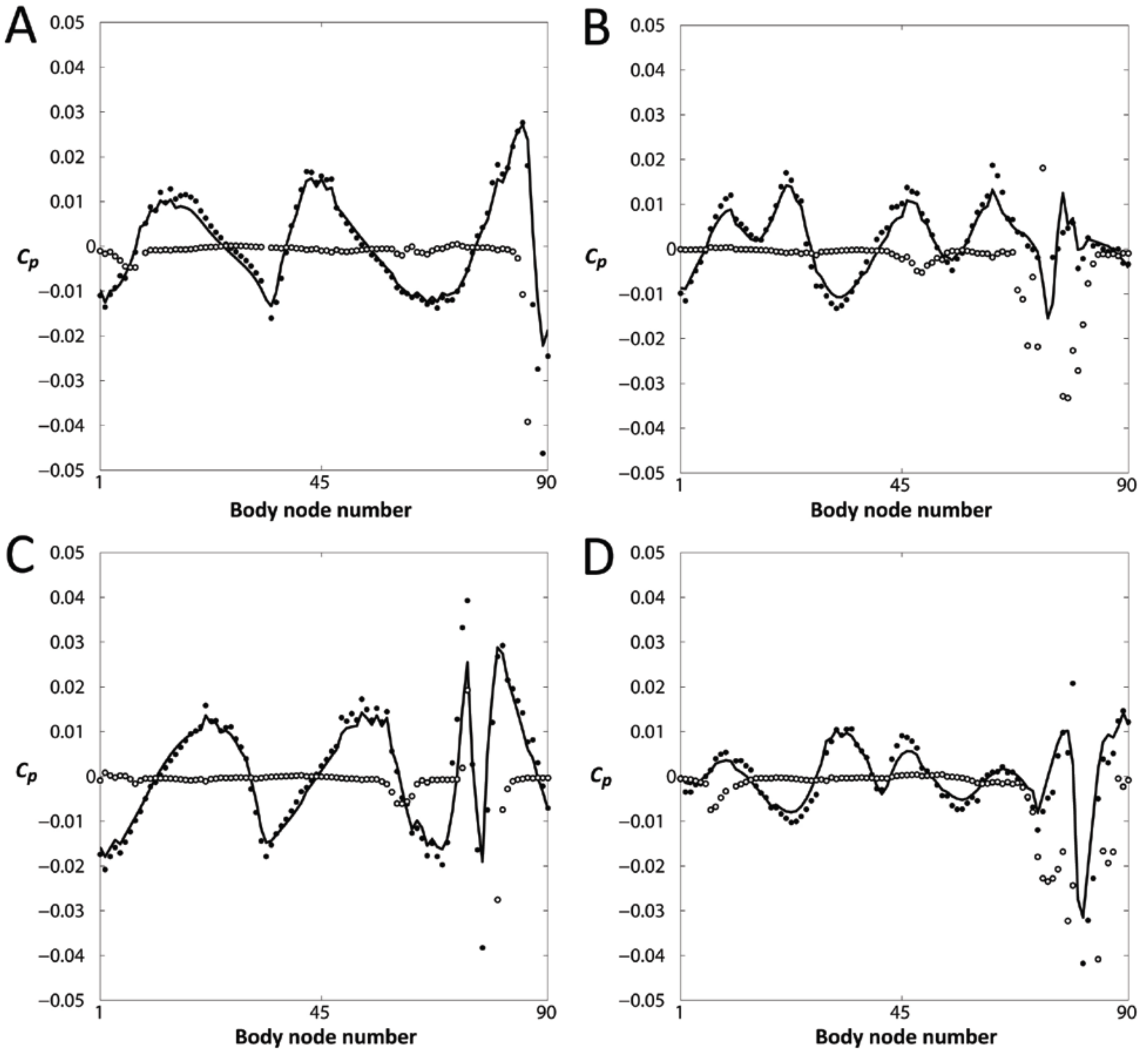

**Figure 4.** Pressure on contour surrounding the self-propelled swimmer and immediately adjacent to the region of undefined pressure, at four instants during the swimming cycle duration ***T*.** Head is at body node number 45; tail is at body nodes 1 and 90. Solid curve, pressure computed from numerical simulation. Closed circles, pressure estimated from unsteady algorithm. Open circles, pressure estimated from quasi-steady algorithm. (A) $t/T$ = 1/4. (B) $t/T$ = 1/2. (C) $t/T$ = 3/4. (D) $t/T$ = 1.

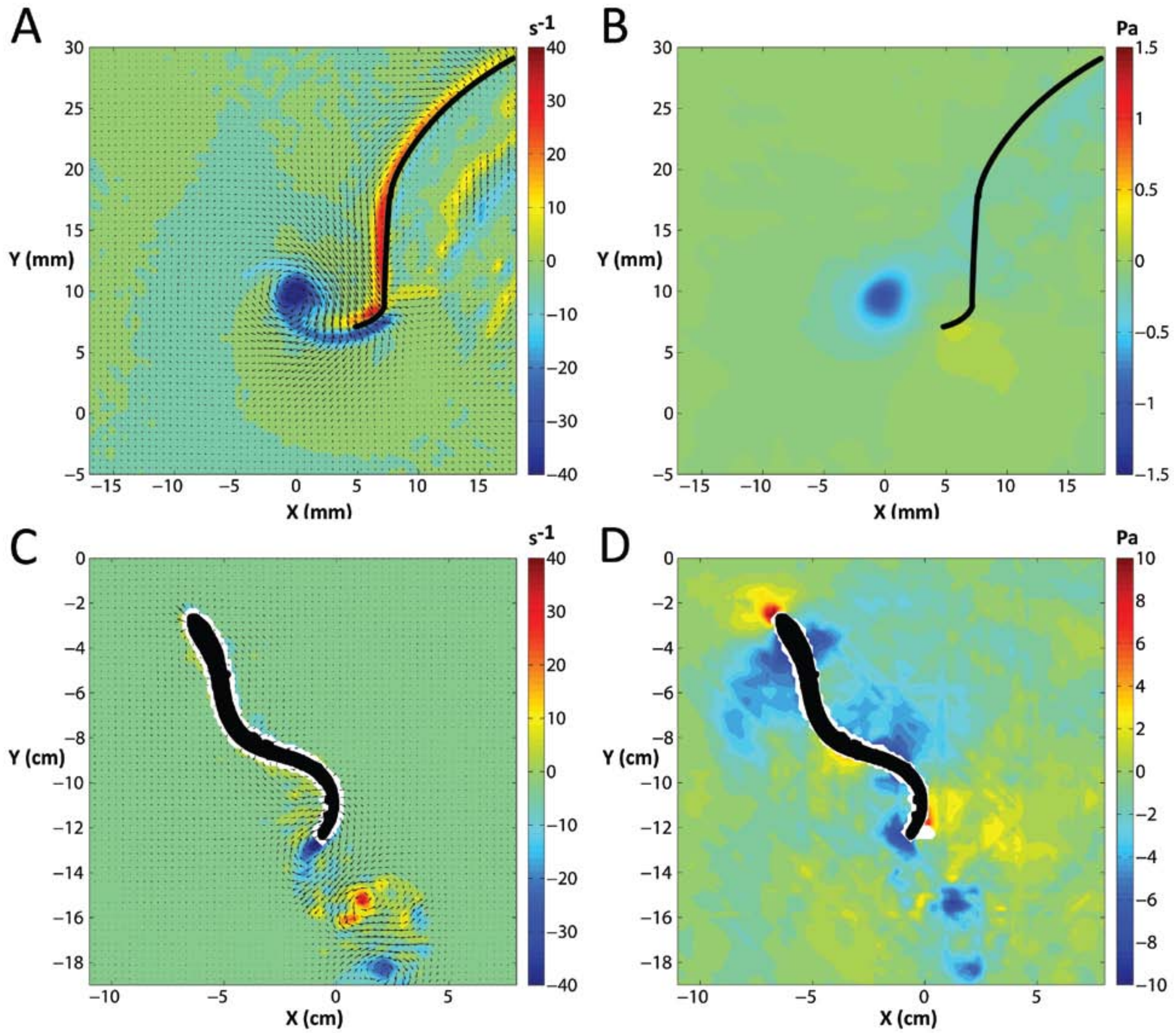

**Figure 5.** (A) Particle image velocimetry (PIV) measurement of a freely-swimming jellyfish medusa. Velocity field is plotted on vorticity contours. Maximum velocity vector is approximately 3 cm s$^{-1}$. Velocity field is plotted at half of full resolution. Left half of exumbrellar surface is indicated by black curve. (B) Pressure field estimated using unsteady algorithm. (C) Particle image velocimetry (PIV) measurement of a freely-swimming lamprey. Velocity field is plotted on vorticity contours. Maximum velocity vector is approximately 11 cm s$^{-1}$. Velocity field is plotted at half of full resolution. Animal body is approximately indicated in black. (D) Pressure field estimated using unsteady algorithm.

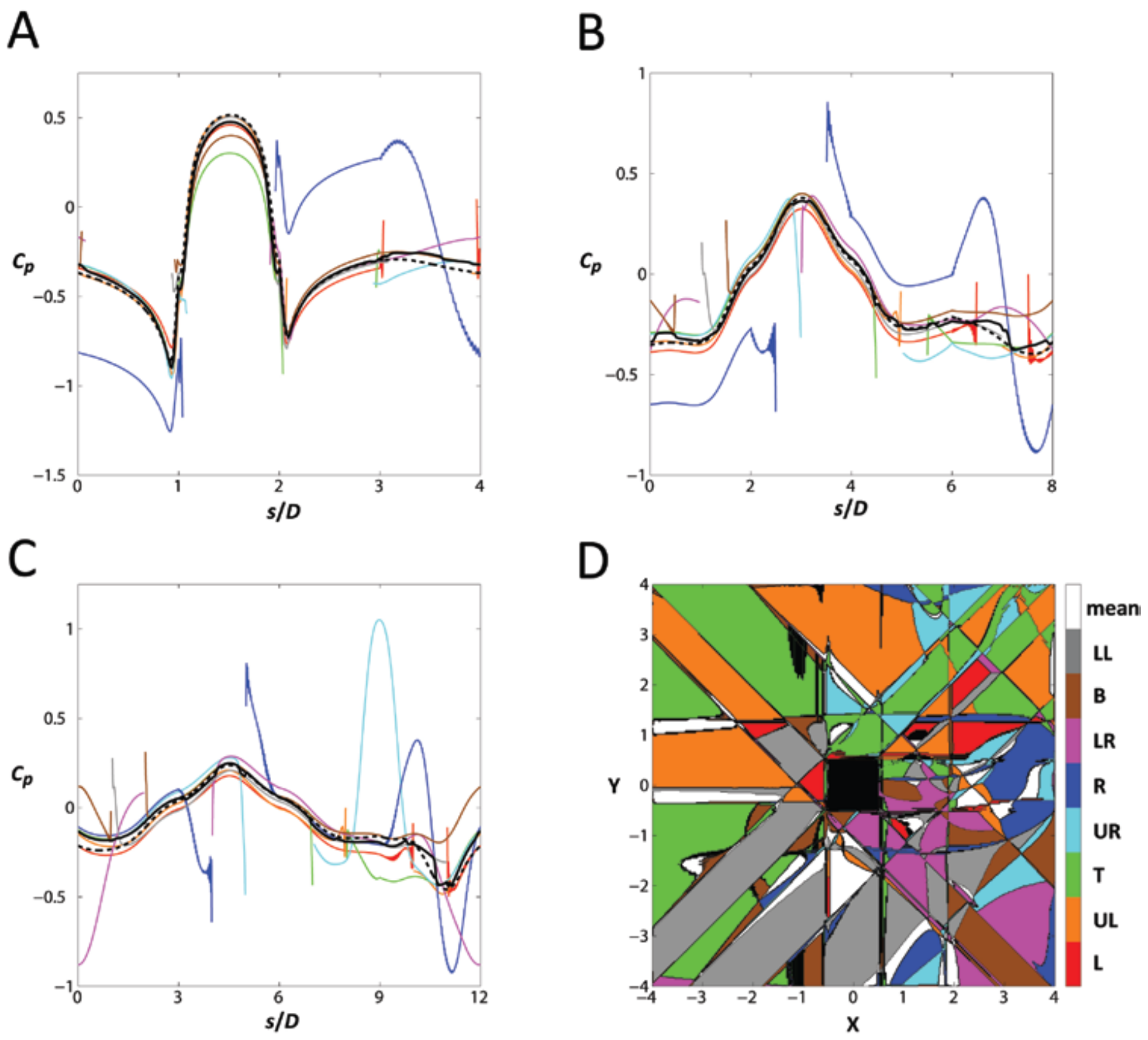

**Figure A1.** (A) Pressure on surface of bluff body estimated using quasi-steady algorithm. Measurement contour is offset by 0.1*D* from the fluid-solid interface. The pressure coefficient $c_p = p/(\rho U^2)$. *s* is the local surface coordinate and increases in the counter-clockwise direction from the upper right corner of the bluff body. Dashed black line, pressure from numerical simulation; solid black line, pressure estimated using quasi-steady algorithm; solid colored lines, pressure estimates based on each family of integration paths. Colors correspond to paths in Fig. 1 and to the legend in panel D. (B, C) Pressure on square contours centered on the bluff body and with side length 2*D* and 3*D*, respectively (e.g. Fig. A2A). *s* is the local surface coordinate and increases in the counter-clockwise direction from the upper right corner of each square contour. The difference in abscissa in panels A-C reflects the different contour perimeters. (D) Contour plot that colors each point in the domain according to the path family that corresponds to the median pressure at that point. To reveal the individual integration path family contributions, a ninth pressure value equal to the mean of the 8 path families is included, so that the median pressure is from either a single integration path family or from the mean.

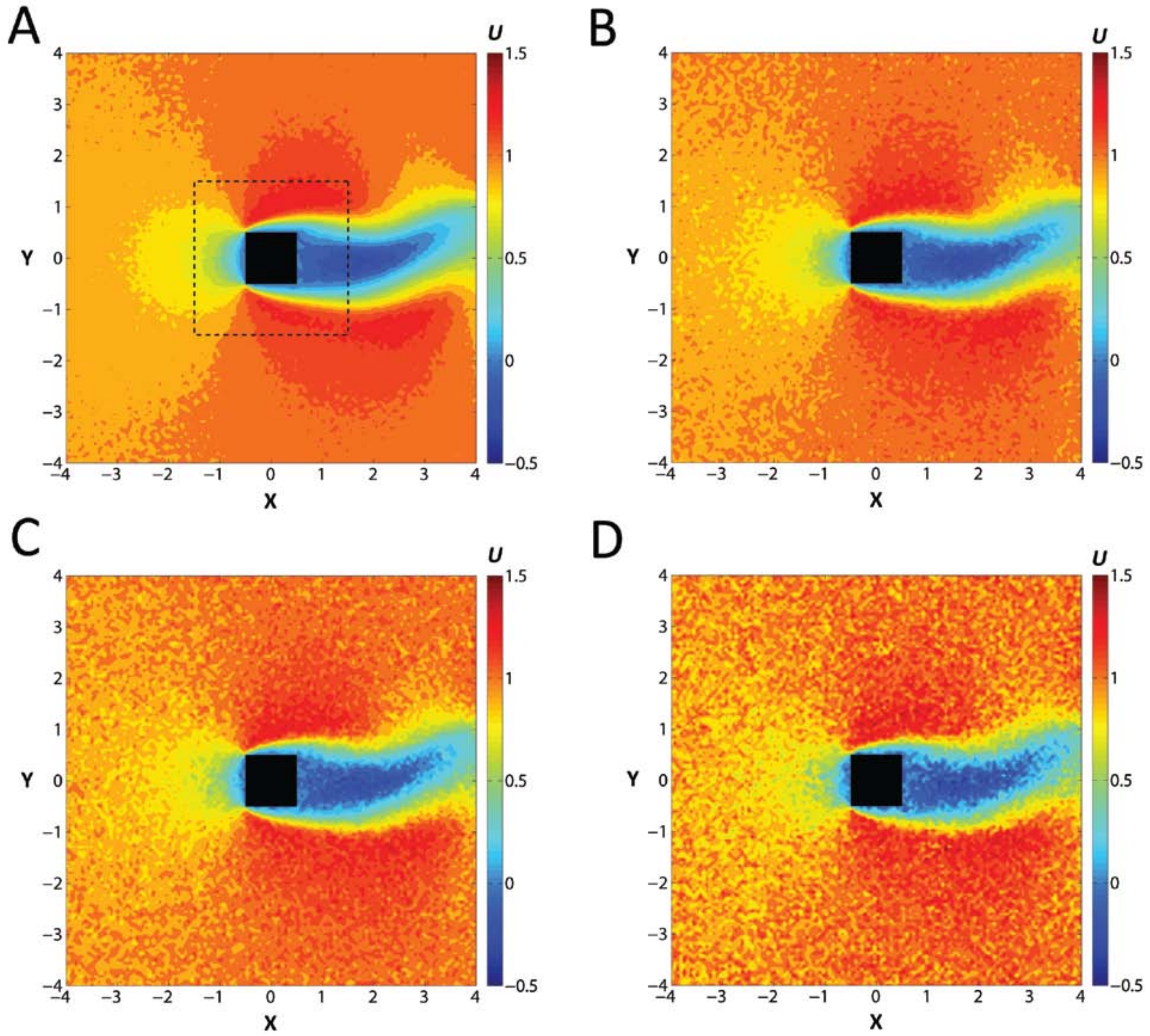

**Figure A2.** Streamwise velocity contours for flow field with Gaussian white noise added to reduce the signal-to-noise ratio (SNR). (A) SNR = 32 dB. (B) SNR = 24 dB. (C) SNR = 20 dB. (D) SNR = 16 dB. Dashed square in panel A indicates contour on which quasi-steady pressure estimates are compared in Fig. A3A.

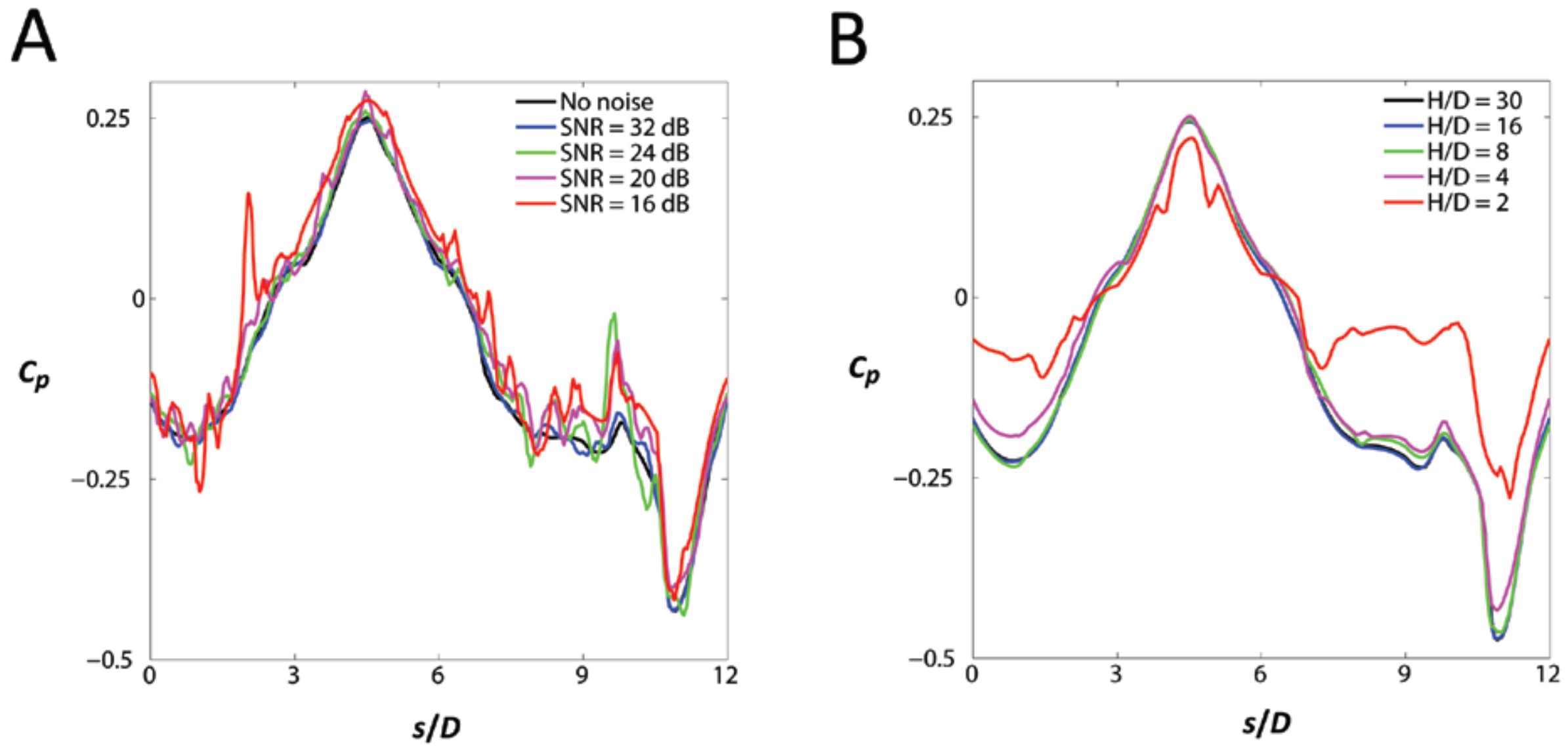

**Figure A3.** (A) Quasi-steady pressure estimate on contour shown in Fig. A2A for varying signal-to-noise ratio. *s* is the local surface coordinate and increases in the counter-clockwise direction from the upper right corner of the square contour. (B) Quasi-steady pressure estimate on contour shown in Fig. A2A for varying measurement domain size. *H* is the half-width of the measurement domain.

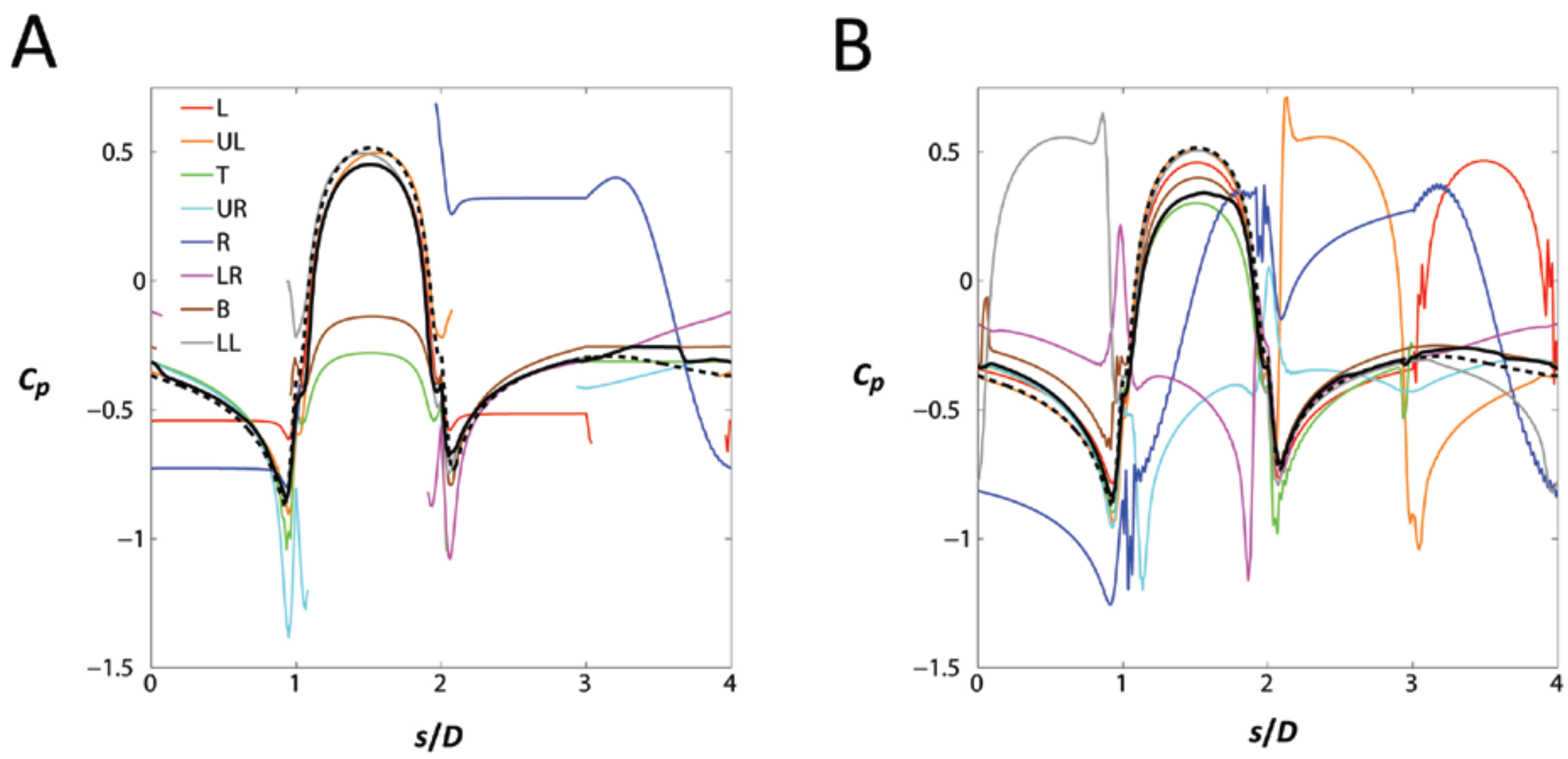

**Figure A4.** (A) Pressure on surface of bluff body estimated using quasi-steady algorithm without viscous term. Measurement contour is offset by 0.1*D* from the fluid-solid interface. *s* is the local surface coordinate and increases in the counter-clockwise direction from the upper right corner of the bluff body. Dashed black line, pressure from numerical simulation; solid black line, pressure estimated using quasi-steady algorithm; solid colored lines, pressure estimates based on each family of integration paths. (B) Pressure on surface of bluff body estimated using quasi-steady algorithm without treatment of fluid-solid interfaces to remove integration paths that pass through the solid body.

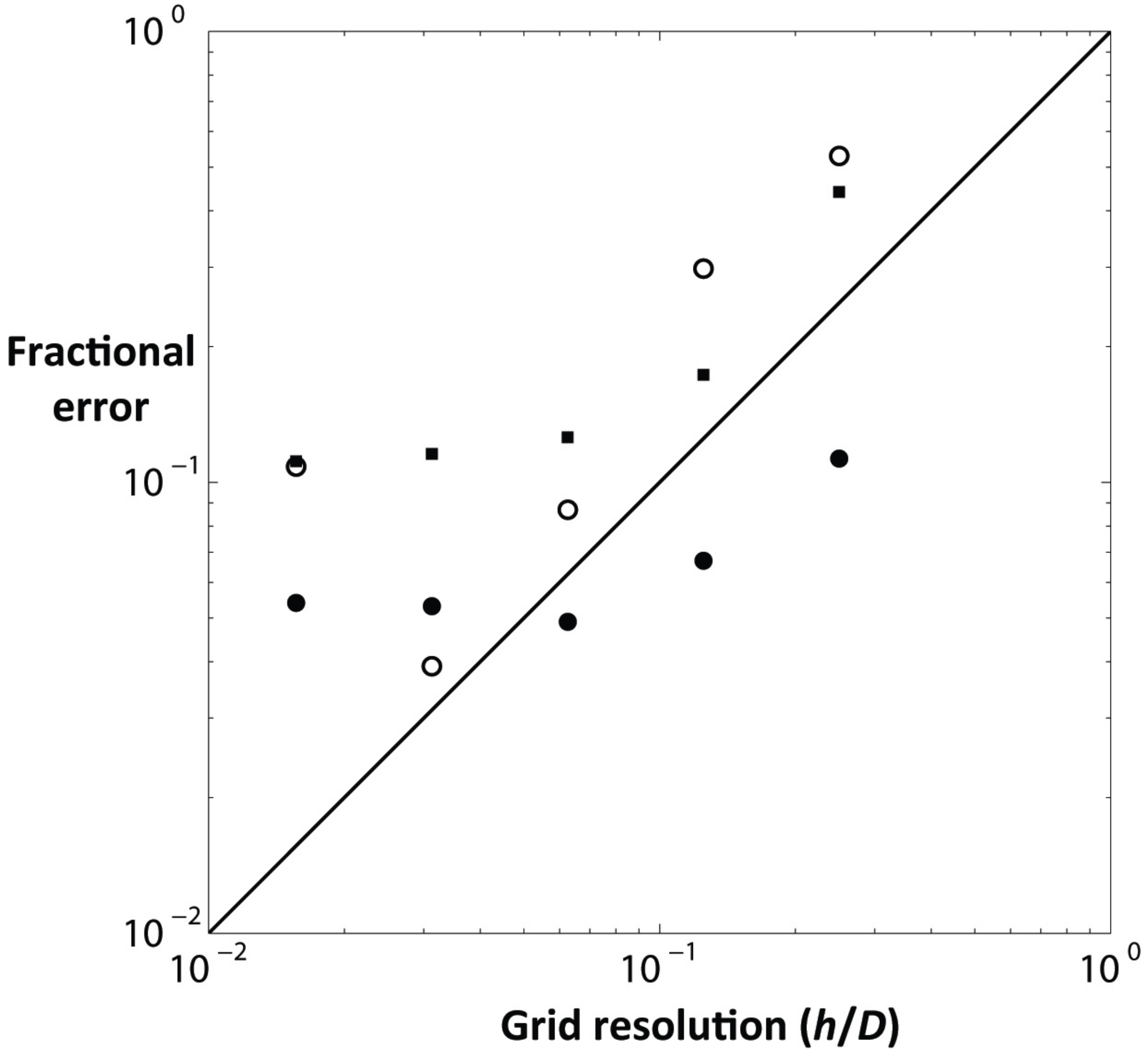

**Figure A5.** Spatial convergence of the algorithm. Log-log plot of the fractional error in instantaneous streamwise (closed circles), instantaneous lateral (open circles), and time-averaged streamwise (closed squares) force coefficients versus grid resolution for numerical simulation of two-dimensional flow past the bluff body. Solid line indicates a slope of 1 corresponding to first-order convergence. Deviation from first-order convergence at small grid resolution is due to a combination of model error and numerical round-off error.

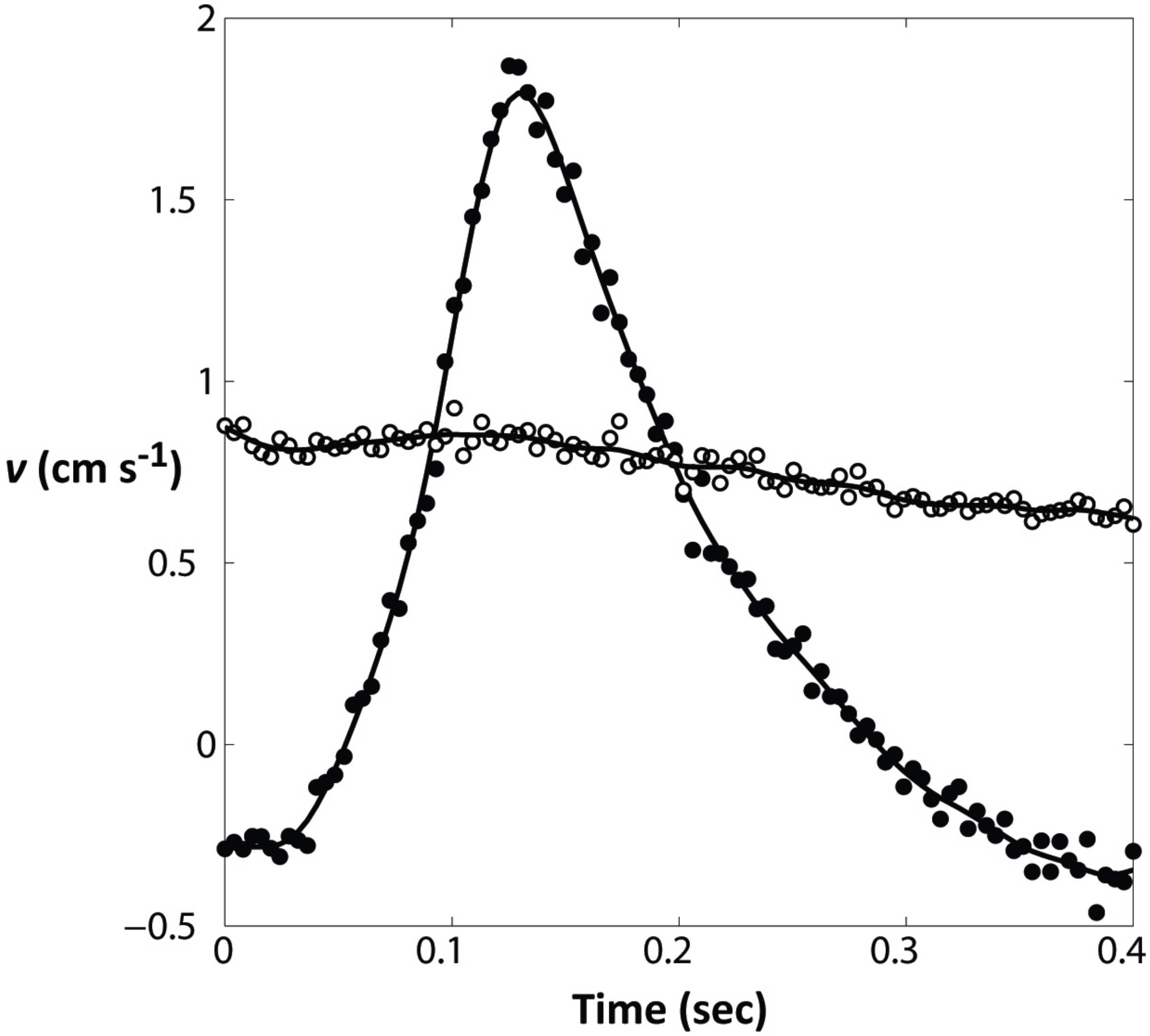

**Figure A6.** Time series of *v* component data at two selected points in the jellyfish PIV data set. Symbols indicate original PIV data at corresponding locations identified in Fig. A7. Solid curves indicate respective smoothing splines.

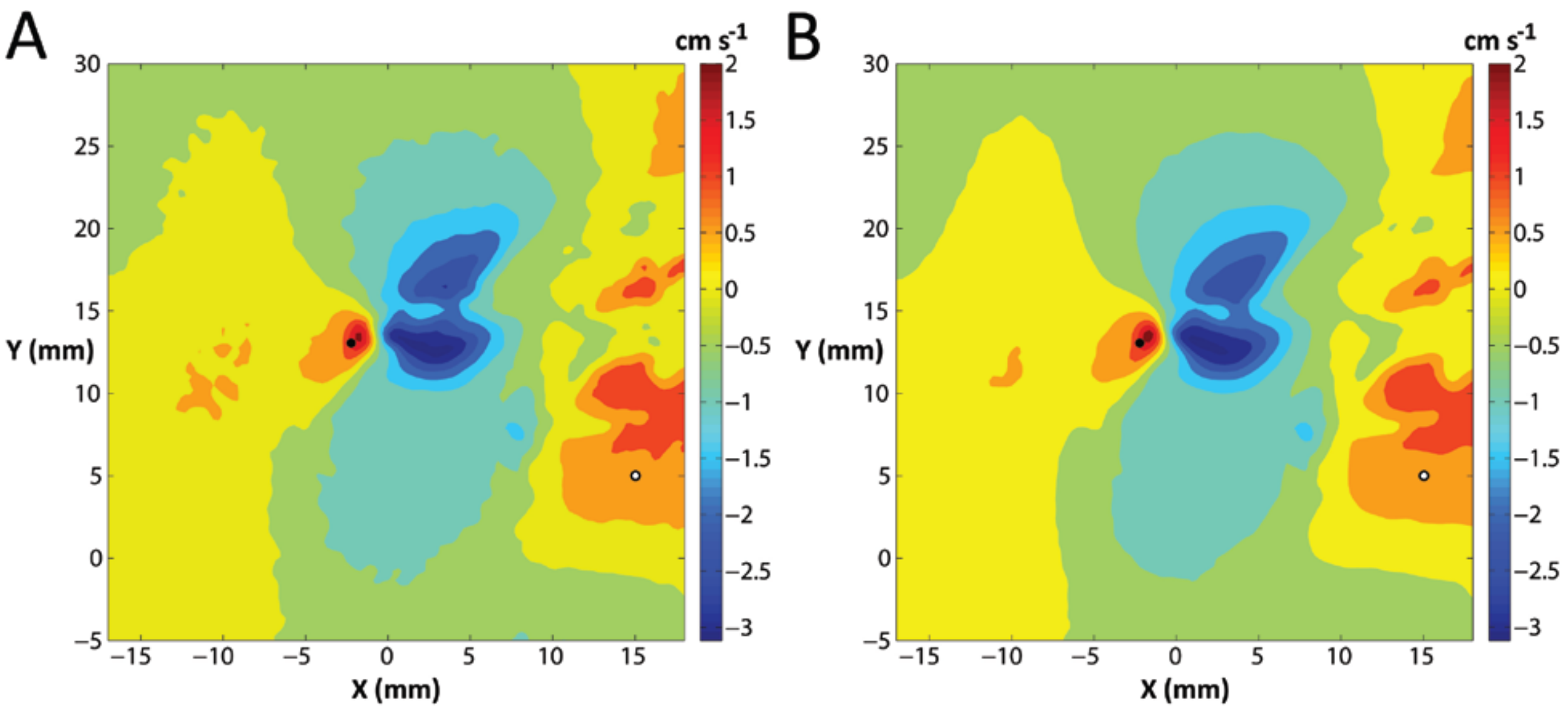

**Figure A7.** (A) Contour plot of *v* component of original velocity measurement during middle of jellyfish bell contraction. (B) Contour plot of *v* component temporal spline-filtered velocity measurement during middle of jellyfish bell contraction. Location of animal is similar to that indicated in Fig. 5A, although earlier in the bell contraction phase. Closed circle near bell margin and open circle in wake indicate locations of temporal profiles in Fig. A6.

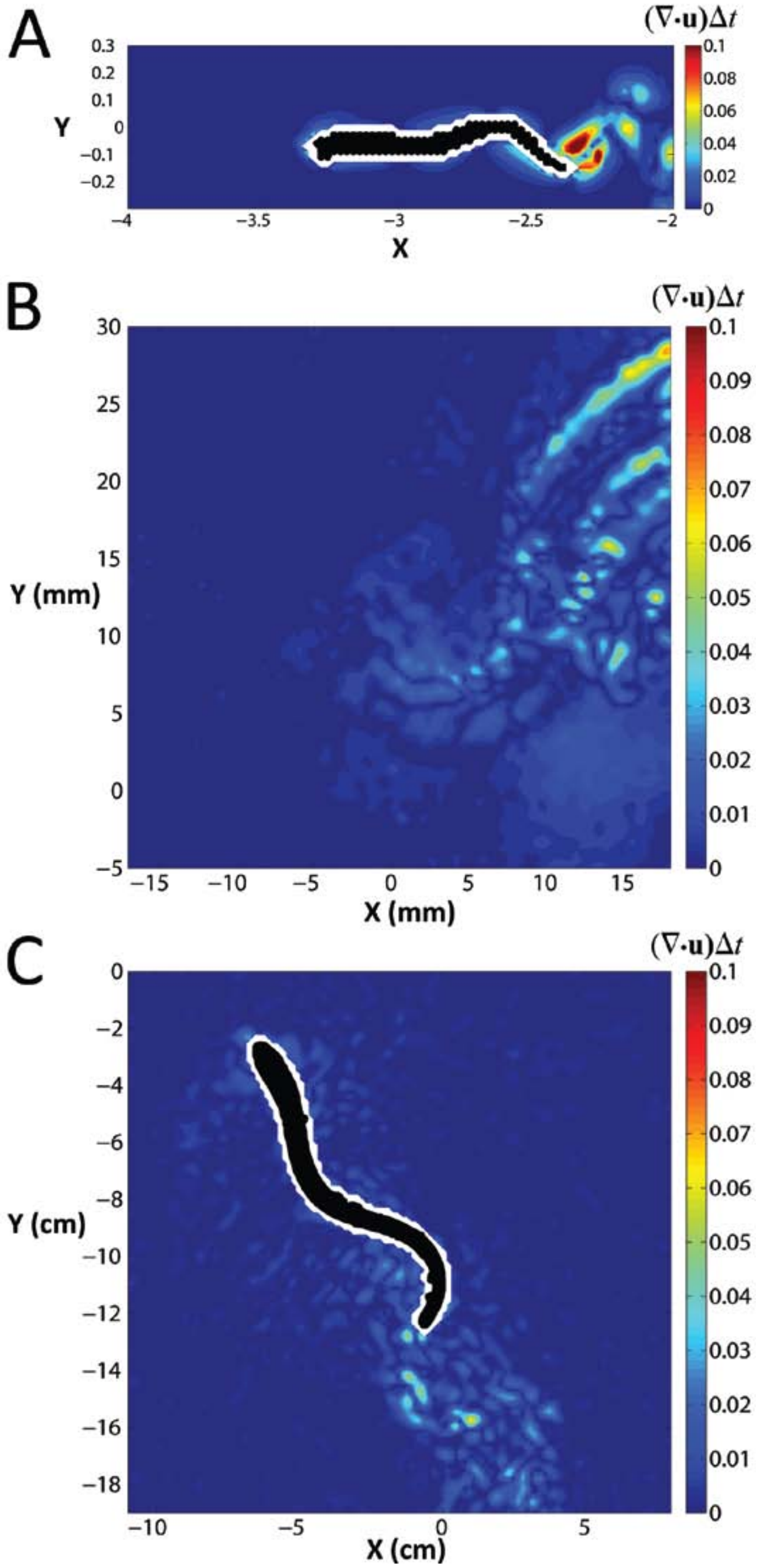

**Figure A8.** Contour plots of normalized two-dimensional divergence for (A) three-dimensional numerical simulation of self-propelled swimmer, (B) PIV measurement of freely-swimming jellyfish (cf. Fig. 5A), (C) PIV measurement of freely-swimming lamprey (cf. Fig. 5C). Dimensional divergence is normalized by multiplying by the time step between sequential velocity fields in each case.

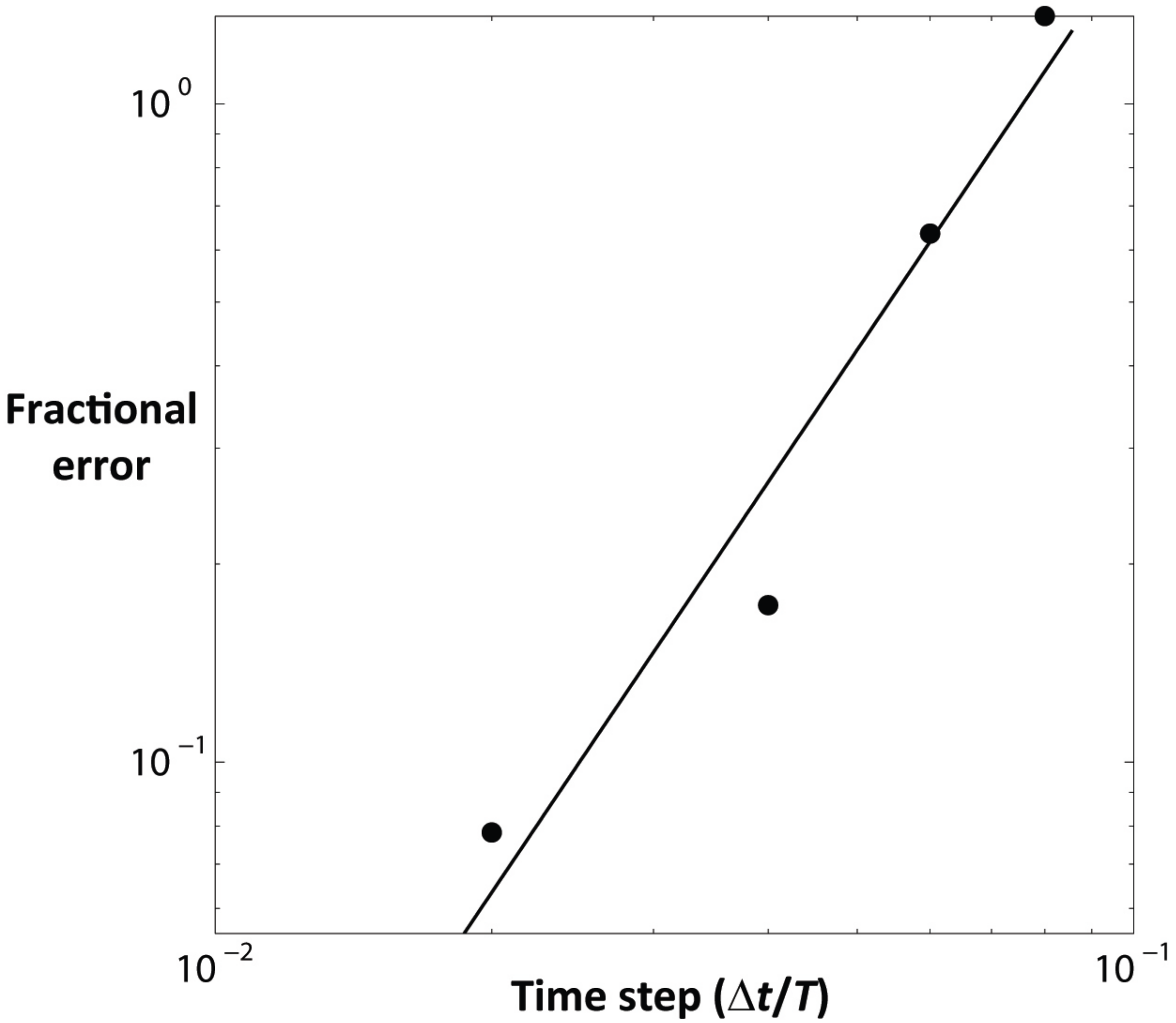

**Figure A9.** Temporal convergence of the algorithm. Log-log plot of the fractional error in pressure at the head of the simulated three-dimensional self-propelled swimmer versus time step between velocity fields (closed circles). Solid line indicates a slope of 2 corresponding to second-order convergence.

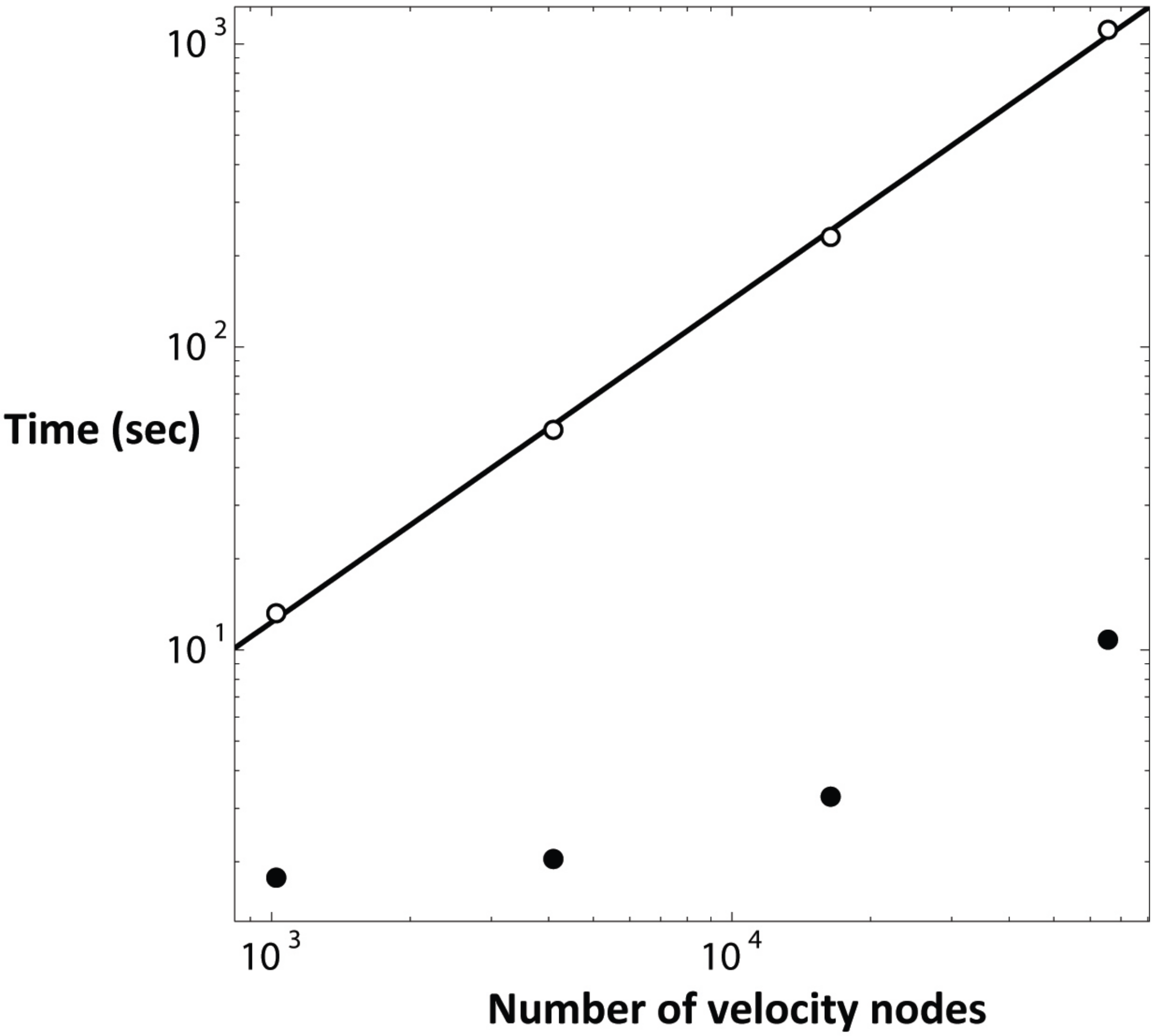

**Figure A10.** Computational cost of the algorithm, as quantified by the time required for a single 3-GHz processor to apply the temporal filter (open circles) and to compute the pressure field (closed circles) for velocity fields from 32x32 to 256x256 nodes. Solid line indicates slope of 1.